           \documentclass[11pt,titlepage,a4paper]{article}
	   \usepackage{amsfonts}

\def\openone{\leavevmode\hbox{\small1\kern-3.8pt\normalsize1}}
\newcommand{\QED}[1][6]{\hbox{\vrule width #1pt height #1pt depth 0pt}}
\newtheorem{Prop}{Proposition}[section]		
\newtheorem{Lem}{Lemma}[section]     		
\newtheorem{Cor}{Corollary}[section] 		

\title{\bfseries	\vspace*{-1.23456789in}
\vspace*{-17ex}
{
\begin{flushright}
	  \textbf{\large LANL xxx E-print archive No.~gr-qc/9709053}\\[12ex]
\end{flushright}
}
{\huge Normal frames and the validity of the equivalence principle}\\
\vspace{10pt}
{\LARGE II. The case along paths} }
\author{
Bozhidar Z. Iliev
\thanks{Department Mathematical Modeling,
Institute for Nuclear Research and Nuclear Energy,
Bulgarian Academy of \mbox{Sciences},
blvd. Tzarigradsko Chauss\'ee 72, 1784 Sofia, Bulgaria}
\thanks{E-mail address: bozho@inrne.acad.bg}
\\\\ \\\\ \\\\ \\\\ \\  \textit{Short title: }
\textbf{\large Normal frames and the equivalence principle: II}
\\\\ \\
\textit{Classification numbers:} 02.40.Hw, 02.40.Ky, 04.20.Cv,
04.50.+h
}
\date{
}

\begin{document}

\renewcommand{\thefootnote}{\fnsymbol{footnote}}
\maketitle			
\renewcommand{\thefootnote}{\arabic{footnote}}

\begin{abstract}

We investigate the validity of the equivalence principle along paths
in gravitational theories based on derivations of the  tensor
 algebra over a differentiable manifold. We prove the existence of
local bases, called normal,  in  which  the components of  the
derivations vanish along arbitrary paths. All such bases are
explicitly  described. The holonomicity of  the normal  bases  is
considered. The results obtained are applied to the important case of
linear  connections and  their relationship with the equivalence
principle is described.
In particular, any gravitational theory based on tensor
derivations which obeys the equivalence principle along all paths,
must be based on a linear connection.

\end{abstract}

\section {Introduction}
\label{I}

A well known classical result is the
existence of local coordinates in which the components of
a symmetric linear  connection~\cite{K&N}  vanish  along  a  smooth
 path without self-intersections~\cite{Schouten/Ricci,Rashevskii}.
For the first time it was observed by
Fermi~\cite{Fermi}  for the Christoffel symbols of a Riemannian
connection
and later it was generalized for arbitrary symmetric linear
connections~\cite[Sec.~25, pp.~64--68]{Eisenhart}. It is natural for
these results to be generalized to the case of nonvanishing torsion.
This is important in connection with the intensive use of nonsymmetric
linear connections~\cite{K&N, Schouten/Ricci} in different physical
theories~\cite{Heyde,MTW}.

	This paper, which is a continuation of~\cite{f-Frames-n+point}
and a revised version of~\cite{f-Bases-path},
investigates the mentioned
problem from the more general viewpoint of arbitrary derivations of the
tensor algebra over a differentiable
manifold~\cite{K&N, Schouten/Ricci}.
In it is proved the existence of special bases (or coordinates),
called \emph{normal},
in which the components of the derivations, as defined below, vanish
along some path.  In particular, our results are valid for linear
connections. The normal bases are explicitly considered and the
question when they are
holonomic or anholonomic~\cite{Schouten/Ricci} is investigated.

	As was pointed out in our previous
work~\cite{f-Frames-n+point},
where the above problems were solved in a neighborhood and at a  point,
the theorem  of  existence  of  normal bases  is  the   right
mathematical background for the consideration of the equivalence
principle (cf.~\cite{MTW,Heyde}).
The results of this paper outline the boundaries  of
validity  of this principle along arbitrary paths in any gravitational
theory  based on derivations.

	The paper is organized as follows. Sec.~2 contains some
preliminary mathematical definitions and results. In Sec.~3 are
investigated problems
concerning normal frames for derivations along arbitrary vector fields.
Sec.~4  and Sec.~5 deal with the same problems but for derivations
along paths and a fixed vector field, respectively.  The results are
specialized for linear connections in Sec.~6. The paper closes Sec.~7
in which connections with the equivalence principle are made.

\section {\bfseries Mathematical preliminaries}
\label{2new}
\setcounter {equation} {0}

	For the explicit mathematical formulation of our problem,  as
well as for reference purposes, in this section we recall some facts
concerning  derivations  of  the tensor algebra  over   a
manifold~\cite{f-Frames-n+point,f-Bases-n+point,K&N}.


Let $D$ be a derivation of the tensor algebra over a manifold
$M$~\cite{K&N}. By~\cite[ proposition 3.3 of chapter I]{K&N} there
exists a unique vector field $X$ and a unique tensor field $S$ of
type $(1,1)$ such that $D=L_{X}+S$.
 Here $L_{X}$ is the Lie derivative along $X$~\cite{K&N}
and $S$ is considered as a derivation of the tensor algebra over
$M$~\cite{K&N}.

\par
If $S$ is a map from the set  of $C^{1}$  vector  fields  into  the
tensor fields of type (1,1) and $S:X\mapsto S_{X}$,  then  the equation
$ D^{S}_{X}=L_{X}+S_{X} $
defines a derivation of the tensor algebra over $M$ for any $C^{1}$
vector field $X$ \cite{K&N}.
Such a derivation will be called an $S$-derivation along $X$ and
denoted for brevity simply by $D_{X}$. An $S$-derivation is a map $D$
such that $D:X\mapsto D_{X}$, where $D_{X}$
is an $S$-derivation along X.

Let $\{E_{i}, i=1,\ldots, n:=\dim (M)\}$ be a (coordinate or
not~\cite{Schouten/Ricci,Lovelock-Rund})
local basis (frame) of vector fields in the tangent bundle to $M$.
It is holonomic (anholonomic) if the vectors
$E_{1}, \ldots , E_{n}$ commute  (do  not
commute)~\cite{Schouten/Ricci,Lovelock-Rund}.
Using the explicit action of $L_{X}$  and $S_{X}$ on  tensor
fields~\cite{K&N} one can easily deduce the explicit form of the local
components of $D_X T$ for any $C^1$ tensor field $T$. In particular,
the \emph{components} $(W_{X})^{i}_{j}$   of $D_{X}$ are defined by
	\begin{equation}	 \label{1}
D_{X}(E_{j})=(W_{X})^{i}_{j}E_{i}.
	\end{equation}
Here and below all Latin indices, perhaps with some
super- or subscripts, run  from  $1$ to $n:=\dim (M)$ and
the usual summation rule on indices repeated  on  different levels is
assumed. It is easily seen that
$(W_{X})^{i}_{j}:=(S_{X})^{i}_{j} - E_{j}(X^{i}) + C^{i}_{kj}X^{k}$
where
$X(f)$ denotes the action of $X=X^{k}E_{k}$ on  the $C^{1}$ scalar
function $f$, as  $X(f):=X^{k}E_{k}(f)$,
and the $C^{i}_{kj}$ define the commutators of the basic
vector fields by $[E_{j},E_{k}]=C^{i}_{jk}E_{i}$.

The change $\{ E_{i}\} \mapsto \{ E^{\prime }_{m}:=A^{i}_{m}E_{i}\} $,
$A:=[ A^{i}_{m}]$  being  a
nondegenerate  matrix  function,  implies  the  transformation  of
$(W_{X})^{i}_{j}$ into (see (\ref{1}))
$(W^{\prime }_{X})^{m}_{l}=(A^{-1})^{m}_{i}A^{j}_{l}$
$(W_{X})^{i}_{j}+(A^{-1})^{m}_{i}X(A^{i}_{l})$. 	
Introducing the  matrices
$W_{X}:=[ (W_{X})^{i}_{j}] $
and
$W^{\prime }_{X}:=[ (W^{\prime }_{X})^{m}_{l}] $
and putting
$X(A):=X^{k}E_{k}(A)=[ X^{k}E_{k}(A^{i}_{m}) ] $, we get
	\begin{equation}	\label{5'}
W^{\prime }_{X}=A^{-1}\{W_{X}A+X(A)\}.
	\end{equation}

	If $\nabla $ is a linear connection with local components
$\Gamma ^{i}_{jk}$
(see, e.g., \cite{K&N,Lovelock-Rund}), then
$\nabla _{X}(E_{j})=(\Gamma ^{i}_{jk}X^{k})E_{i}$~\cite{K&N}.
Hence,  we see from (\ref{1}) that
$D_{X}$ is a covariant differentiation along $X$ iff
	\begin{equation}	\label{7}
(W_{X})^{i}_{j}=\Gamma ^{i}_{jk}X^{k}
	\end{equation}
for some functions $\Gamma ^{i}_{jk}$.

Let $D$ be an S-derivation and $X$ and $Y$ be vector fields.
The \emph{torsion operator} $T^{D}$ of $D$ is defined as
		\begin{equation}   \label{8}
T^{D}(X,Y):=D_{X}Y-D_{Y}X-[X,Y].
		\end{equation}
The S-derivation $D$ is \emph{torsion free} if $T^{D}=0$
(cf. \cite{K&N}).

For a linear connection $\nabla $, due to (\ref{7}), we have
%
$(T^{\nabla }(X,Y))^{i}     =  T^{i}_{kl}X^{k}Y^{l}$
%
where
 $ T^{i}_{kl}
    := - (\Gamma ^{i}_{kl} - \Gamma ^{i}_{lk}) - C^{i}_{kl}$
are the components  of the torsion tensor of $\nabla $~\cite{K&N}.


Mathematically the task of this work is to investigate the  problem
of when along a given path $\gamma :J\to M, J$ being a  real
interval, there exist special frames
$\{E_{i}^{\prime }\}$, called
\emph{normal},
in which the components $W_{X}^{\prime}$ of an S-derivation $D$ along
some or all vector fields $X$ vanish. In  other words, we are going to
solve equation~(\ref{5'}) with respect to $A$ under certain conditions,
which will be specified below. Physically the solution of this problem
corresponds  to  the investigation of the validity  of the equivalence
principle along paths.

\section {\bfseries Derivations along arbitrary vector fields}
\label{II}
\setcounter {equation} {0}

In this section we investigate  the  problem of
existence and some properties of special bases $\{E_i^{\prime }\}$ in
which the components of a given S-derivation $D_{X}$ along an
\emph{arbitrary vector field}  $X$ vanish along a path
$\gamma:J\to M$, with $J$ being  an $\mathbb{R}$-interval.
Such bases or frames will be called
\emph{normal along} $\gamma$. Note that $\{E_i^{\prime }\}$ are
supposed to be defined \emph{in  a  neighborhood}  of $\gamma (J)$,
while the components of $D$ vanish \emph{on} $\gamma (J)$.

	The S-derivation $D$ is \emph{linear along} $\gamma$ if for
all $X$ in some (and hence in any) basis $\{E_{i}\}$,
we have (cf.~(\ref{7}))
	\begin{equation}	\label{12}
W_{X}(\gamma (s))=\Gamma _{k}(\gamma (s))X^{k}(\gamma (s))
	\end{equation}
for some matrix  functions $\Gamma _{k}$ defined on $\gamma (J)$.
This means that~(\ref{7}) is valid for $x\in \gamma (J)$, but it may
not be valid for $x\not\in \gamma (J)$. Evidently, a linear connection
is a linear derivation along any path $\gamma$.

	\begin{Prop}	\label{Prop1}
 An S-derivation $D$  is
linear along a path $\gamma :J\to M$ if and only if along $\gamma $
there exists a normal frame for $D$, i.e. one  in  which the components
of $D_X$ along every vector field  $X$ vanish along $\gamma $ (that is,
on $\gamma (J)$).
	\end{Prop}

\textit{Proof.} Let the derivation $D$ be  linear along $\gamma $,
i.e.~(\ref{12}) is valid. Let us at first assume that $\gamma $ is
without
self-intersections and that $\gamma (J)$ is contained in  only  one
coordinate neighborhood $U$ in which some local coordinate basis
$\{E_{i}=\partial/\partial x^{i}\}$ is fixed.

Due to~(\ref{5'}) we have  to  prove  the  existence  of  a  matrix
$A=[  A^{i}_{j } ] $
%
%
for which in the basis
$\{E_{j}^{\prime }=A^{i}_{j}E_{i}\}$ the equality
$ W^\prime_{X}(\gamma (s))=0$ is fulfilled for every $X=X^{k}E_{k}$.
Substituting~(\ref{12}) into~(\ref{5'}), we see that the last equation
is equivalent to
	\begin{equation}	\label{13}
\Gamma _{k}(\gamma (s))A(\gamma (s)) +
E_{k}(A)\mid _{\gamma (s)}=0, \quad
E_{k}=\partial /\partial x^{k}.
	\end{equation}
The general solution  of  this  equation  can  be  constructed  as
follows.

Let $V:=J\times \cdots \times J$, where $J$ is taken $n-1$ times.
Let us fix a one-to-one $C^{1}$ map $\eta :J\times V\to M$ such  that
$\eta (\cdot ,\mathbf{t}_{0})=\gamma $  for  some
fixed
$\mathbf{t}_{0}\in V$, i.e. $\eta(s,\mathbf{t}_{0})=\gamma(s), s\in J$.
This is  possible  iff $\gamma $  is without self-intersections. In
$U\bigcap \eta (J,V)$  we  introduce coordinates $\{x^{i}\}$ by putting
$(x^{1}(\eta (s,\mathbf{t})),\ldots ,x^{n}(\eta (s,\mathbf{t}))) =
(s,\mathbf{t})$, $s\in J, \mathbf{t}\in  V$.
This, again, is possible iff $\gamma $ is without self-intersections.

If we expand $A(\eta (s,\mathbf{t}))$ into a power series with respect
to $(\mathbf{t}-\mathbf{t}_{0})$,
we find the general solution of~(\ref{13}) in the form
	\begin{eqnarray}	\nonumber
A(\eta (s,\mathbf{t})) =
\left\{
\openone - \sum^{n}_{k=2}\Gamma _{k}(\gamma (s))
[x^{k}(\eta (s,\mathbf{t}))-x^{k}(\eta (s,\mathbf{t}_{0}))]
\right\} &\times&
\\  \nonumber \times \
Y(s,s_{0};-\Gamma _{1}\circ \gamma )B(s_{0},\mathbf{t}_{0};\eta ) &+&
\\ \label{14} + \
B_{kl}(s,\mathbf{t};\eta )
[x^{k}(\eta (s,\mathbf{t}))-x^{k}(\eta (s,\mathbf{t}_{0}))]
[x^{l}(\eta (s,\mathbf{t}))-x^{l}(\eta (s,\mathbf{t}_{0}))]. & &
	\end{eqnarray}
Here: ${\openone}$ is the unit matrix, $s_{0}\in J$ is fixed, $B$ is
any nondegenerate matrix function of its arguments, the matrix
functions $B_{kl}$ are such that they and their first derivatives
are bounded when
$\mathbf{t}\to \mathbf{t}_{0}$, and $Y=Y(s,s_{0};Z)$, with $Z$ being a
continuous matrix function of $s$, is the unique solution  of
the matrix initial-value problem~\cite[ch. IV, \S1]{Hartman}
	\begin{equation}	\label{15}
{dY\over ds} =ZY, \quad \left.Y\right|_{s=s_{0}}={\openone}, \quad
Y=Y(s,s_{0};Z).
	\end{equation}
Hence, a matrix $A$, and, consequently, a basis $\{E_{i}^\prime\}$
with  the needed property exist.

If $\gamma (J)$ does not lie in  only  one  coordinate  neighborhood,
then, by means of the above described method, we  can  obtain  local
normal frames in different coordinate  neighborhoods
which form a neighborhood of $\gamma (J)$. From these local normal
bases we can construct a global normal basis along $\gamma $.
  Generally this frame will not be continuous in the regions  of
intersection of two or more coordinate neighborhoods.
For example, suppose for
some $\gamma $ there doesn't exist one coordinate neighborhood
containing $\gamma (J)$
but there  are  two coordinate  neighborhoods
$U^\prime $  and $U^{\prime\prime}$
such that
$\gamma (J)\subset U^\prime \bigcup U^{\prime\prime}$.
Then in $U^\prime $ and $U^{\prime\prime}$  there  are  (see  above)
normal bases $\{E_{i}^\prime\}$ and $\{E_{i}^{\prime\prime}\}$ along
$\gamma$ for $D_{X}$ for  every $X$. So, a global normal basis
$\{E^{0}_{i}\}$ in $U^\prime \bigcup U^{\prime\prime}$
can be obtained  by  putting
$\left.E^{0}_{i}\right|_{x} =
\left.E_{i}^{\prime }\right|_{x}$
for $x\in U^\prime $  and
$\left.E^{0}_{i}\right|_{x} =
\left.E_{i}^{\prime\prime}\right|_{x}$ for
$x\in U^{\prime\prime}\backslash U^\prime $
(note that $U^{\prime\prime}\bigcap U^\prime $ is not empty  as
$\gamma $ is $C^{1}$ path).

Analogously, if $\gamma $ has self-intersections, then on  any  `part'
of $\gamma $ without self-intersections there exist local
normal frames. From these frames can be
constructed a global normal one along $\gamma$.
(At the points of self-intersections of $\gamma $ we can arbitrary fix
these bases to be the ones obtained  above for some fixed part of
$\gamma $ without self-intersections.)

Consequently, if $D$ is linear  along $\gamma$, then in  a
neighborhood of $\gamma (J)$ a  basis $\{E_{i}^{\prime }\}$ which is
normal along $\gamma $ exists for $D_{X}$ for every vector field $X$.

Conversely, let us assume the existence of a
frame $\{E_{i}^{\prime }\}$ which is normal along $\gamma$, i.e.
$W^\prime_{X}=0$ for every  X.  Fixing  some  basis $\{E_{i}\}$  such
that $E_{j}^{\prime }=A^{i}_{j}E_{i}$,  from~(\ref{5'})  we  find
$\left. (W_{X}A+X(A))\right|_{\gamma (s)}=0$.  Consequently
 $W_{X}(\gamma (s)) = -\left. [(X(A))A^{-1}]\right|_{\gamma (s)}$
which means that the equation~(\ref{12}) holds  for
$\Gamma _{k}(\gamma (s))=
- \left. [(E_{k}(A))A^{-1}]\right|_{\gamma (s)}$.~\QED

	\begin{Prop}	\label{Prop2}
All frames which are normal along a path $\gamma $  for an
S-derivation, if any,  are connected along
$\gamma $ by linear transformations  whose coefficients are
such that the action of the  vectors  from  these bases on them vanish
along $\gamma$  (i.e. on $\gamma (J)$).
	\end{Prop}

\textit{Proof.} If $\{E_{i}\}$  and $\{E_{i}^{\prime }\}$  are  normal
frames, then we have
$W^\prime_{X}(\gamma (s))=W_{X}(\gamma (s))=0$.
So, from~(\ref{5'}) follows
$\left. X(A)\right|_{\gamma (s)}=0$ for  every  vector field
$X=X^{k}E_{k}$,
i.e.  $\left. E_{k}(A)\right|_{\gamma (s)}=0$.~\QED

	\begin{Prop}	\label{Prop3}
If along a path $\gamma :J\to M$  there is a local \textit{holonomic}
(on $\gamma (J)$) normal frame for some S-derivation $D$,
then $D$ is torsion free on $\gamma (J)$. Conversely, if $D$  is
torsion free on $\gamma (J)$ and a smooth ($C^1$) normal frame for $D$
along $\gamma $ exists, then all frames which are normal for $D$
are holonomic along $\gamma $.
	\end{Prop}

\textbf{Remark.} In the second part of this proposition we demand the
frames to be smooth. This is necessary as any holonomic basis
is such. Besides, as we saw in the proof of proposition~\ref{Prop1}, if
$\gamma (J)$ is not contained in only one coordinate neighborhood or if
$\gamma $ has self-intersection, then, generally, along $\gamma $ there
does  not exist a continuous, even anholonomic, basis with the needed
property. But on any piece of $\gamma $ without self-intersection which
lies in only one coordinate neighborhood a continuous, but maybe
anholonomic, normal basis exists.

\textit{Proof:}
If $\{E_{i}^{\prime }\}$ is a normal basis, i.e.
$W^\prime_{X}(\gamma (s))=0$  for  every $X$  and $s\in J$,  then,
using~(\ref{8}), we  find
$\left. T^{D}(E_{i}^{\prime },E_{j}^{\prime }) \right|_{\gamma (s)}=
- \left. [E_{i}^{\prime }, E_{j}^{\prime }] \right|_{\gamma (s)}$.
Consequently $\{E_{i}^{\prime }\}$
is holonomic  at $\gamma (s)$, i.e.
$\left. [E_{i}^{\prime },E_{j}^{\prime }] \right|_{\gamma (s)}=0,$
iff
\(
0=\left. T^{D}(X,Y) \right|_{\gamma (s)}=
X^{{\prime}i}(\gamma (s)) Y^{{\prime}j}(\gamma (s))
(T^{D}(E_{i}^{\prime },\left. E_{j}^{\prime }) \right|_{\gamma (s)})
\)
(see proposition~\ref{Prop1} and~(\ref{12}))
for every vector fields $X$ and $Y$, which is equivalent to
$\left. T^{D}\right|_{\gamma (J)}=0.$

Conversely, let $\left.T^{D}\right|_{\gamma (J)}=0$. We have  to
prove  that  any basis
$\{E_{i}^{\prime }\}$ along $\gamma $ in which
$W^\prime_{X}(\gamma (s))=0$
is holonomic at $\gamma (s)$, $s\in J$.
The holonomicity at $\gamma (s)$  means
\(
0=\left. [E_{i}^{\prime },E_{j}^{\prime }] \right|_{\gamma (s)} =
\left.\left\{
 (A^{-1})^{l }_{k} \left(
  			E_{i}^{\prime }(A^{k}_{j }) -
  			E_{j}^{\prime }(A^{k}_{i })
		   \right) E_{l}^{\prime }
\right\}\right|_{\gamma (s)}.
\)
But (see  proposition~\ref{Prop1})  the  existence  of
$\{E_{i}^{\prime }\}$ is equivalent to
$W_{X}(\gamma (s))=(\Gamma _{k} \left. X^{k}) \right|_{\gamma (s)}$
for every X.
These two facts, combined with~(\ref{8}), show that
$\left. (\Gamma _{k})^{i}_{ j} \right|_{\gamma (s)} =
\left. (\Gamma _{j})^{i}_{ k} \right|_{\gamma (s)}$.
Using this and
\(
\left. (\Gamma _{k}A+\partial A/\partial x^{k}) \right|_{\gamma (s)}=0
\)
(see the proof of proposition~\ref{Prop1}),  we find
\(
\left. E_{j}^{\prime }(A^{k}_{i }) \right|_{\gamma (s)} =
- \left.  A^{l}_{j }A^{m}_{i }
(\Gamma _{l})^{k}_{m } \right|_{\gamma (s)} = \left. E_{i}^{\prime }
(A^{k}_{j }) \right|_{\gamma (s)}.
\)
  Therefore
$\left. [E_{i}^{\prime },E_{j}^{\prime }] \right|_{\gamma (s)}=0$
(see above), i.e $\{E_{i}^{\prime }\}$ is a holonomic normal frame on
$\gamma(J)$.~\QED

It can be proved (see lemma~\ref{Lem7} below) that  for  any  path
$\gamma :J\to M$ every frame $\{E^{\gamma }_{i}\}$
\emph{defined only on}
$\gamma (J)$ can locally be extended to a
\emph{holonomic} frame $\{E^{h}_{i}\}$
\emph{defined in a neighborhood of} $\gamma (J)$  and  such
that
$\left. E^{h}_{i}\right|_{\gamma (J)}=E^{\gamma }_{i}$.
In particular,
this is true for  the  restriction
\(
{^\prime \!}E^{\gamma }_{i} = \left.E_{i}^{\prime }\right|_{\gamma (J)}
\)
of the normal bases $\{E_{i}^{\prime }\}$ considered above.  But  in
the general case, the extended holonomic bases
$\{ {^\prime \!}E^{h}_{i}\}$ will not have
the special property that $\{E_{i}^{\prime }\}$ has.

\section {\bfseries Derivations along paths}
\label{III}
\setcounter {equation} {0}

Let $\gamma :J\to M, J$ being an $\mathbb{R}$-interval, be a $C^{1}$
path and $X$  be a $C^{1}$ vector field defined in a neighborhood of
$\gamma (J)$ in such  a  way that on $\gamma (J)$ it reduces to the
tangent vector field $\dot{\gamma }$,  i.e.
$\left.X\right|_{\gamma (s)}=\dot{\gamma }(s)$, $s\in $J.
We call the restriction on $\gamma (J)$ of an S-derivation $D_{X}$
along $X$ (S-)derivation along $\gamma $ and  denote  it  by
${\mathcal D}^{\gamma }$.
Of course, ${\mathcal D}^{\gamma }$ generally depends
on the values of $X$ outside $\gamma (J)$, but, as this dependence is
insignificant for the following, it will not be written explicitly. So,
if $T$ is a $C^{1}$ tensor field in  a  neighborhood of $\gamma (J)$,
then
	\begin{equation}	\label{16}
 ({\mathcal D}^{\gamma }T)(\gamma (s)) := {\mathcal D}^{\gamma}_{s}T :=
\left.(D_{X}T)\right|_{\gamma (s)}, \quad
\left. X\right|_{\gamma (s)}=\dot{\gamma }(s).
	\end{equation}
It is easily seen that ${\mathcal D}^{\gamma }_{s}T$ depends only on
the values of $\left.T\right|_{x}$ for $x\in \gamma (J)$, but not on
the ones for
$x\not\in \gamma (J)$. The operator ${\mathcal D}^{\gamma }$ is  a
generalization of the usual covariant differentiation along curves
(see~\cite{Schouten/Ricci,Rashevskii,MTW} or Sect.~\ref{V}).

When restricted to $\gamma (J)$, the components of $D_{X}$ will be
called components of ${\mathcal D}^{\gamma }$.

	\begin{Prop}	\label{Prop4}
Along any $C^{1}$ path $\gamma :J\to M$ there exists a basis
$\{E_{i}^{\prime }\}$ in which the components of a given S-derivation
${\mathcal D}^{\gamma }$ along $\gamma $  vanish on $\gamma(J)$.
	\end{Prop}

\textit{Proof.} Let us fix a  basis  $\{E_{i}\}$ in  a  neighborhood
of $\gamma (J)$. We have to prove the existence of  a transformation
$\{E_{i}\} \to  \{E_{j}^{\prime } = A^{i}_{j }E_{i}\}$ such that
$\left. W_{X}^\prime\right|_{\gamma (J)}=0$. By~$(\ref{5'})$ this is
equivalent  to the existence of a matrix function
$A=[ A^{i}_{j } ] $ satisfying  along $\gamma $
the equation
\(
\left.\left( A^{-1}(W_{X}A+X(A)) \right)\right|_{\gamma (J)}=0, s\in J,
\)
or
	\begin{equation}	\label{17}
\left. \dot{\gamma }(A)\right|_{\gamma (s)}
 \equiv
{dA(\gamma (s))\over ds} = - W_{X}(\gamma (s))A(\gamma (s))
\end{equation}
as $\left.X\right|_{\gamma (s)}=\dot{\gamma }(s)$.
	The general solution of this equation with respect to $A$ is
	\begin{equation}	\label{18}
\mathbf{ A}(s;\gamma )=Y(s,s_{0};-W_{X}\circ \gamma )B(\gamma ),
	\end{equation}
where $Y$ is the unique solution of the initial-value
problem~(\ref{15}),
$s_{0}\in J$ is fixed, and $B(\gamma )$ is a nondegenerate matrix
function of $\gamma $.

	Let $A$  be  any  matrix  function   with  the  property
$\left.A(x)\right|_{x=(\gamma (s))} = \mathbf{A}(s;\gamma )$
for some $s_{0}$ and $B$. (E.g., using the notation  of  the proof of
proposition~\ref{Prop1}, in any coordinate neighborhood in which
$\gamma $ is without self-intersections, we can put
\(A(\eta (s,\mathbf{t})) =
Y(s,s_{0};-W_{X}\circ \gamma )B(s_{0},\mathbf{t}_{0},\mathbf{t};\gamma)
\)
for a fixed nondegenerate matrix function $B$.) Then it is easily seen
that $A$ carries out the needed  transformation.  Hence, the  basis
$\{ E_{j}^{\prime }=A^{i}_{j }E_{i} \}$ is the one looked for.~\QED

	\begin{Prop}	\label{Prop5}
The normal frames along $\gamma :J\to M$ for  ${\mathcal D}^{\gamma}$
are connected by linear transformations whose coefficients on
$\gamma(J)$ are constant or  may depend only on $\gamma $.
	\end{Prop}

\textit{Proof.} If $\{E_{i}\}$ and $\{E_{i}^{\prime }\}$
are  normal bases, then
$W_{X}(\gamma (s))=W_{X}^\prime(\gamma (s))=0$,
$\left.X\right|_{\gamma (s)}=\dot{\gamma }(s)$
So, from~(\ref{5'}) follows
$\left.\dot{\gamma }(A)\right|_{\gamma (s)}=dA(\gamma (s))/ds=0,$ i.e.
$A(\gamma(s))$ is a constant or depends only on the map $\gamma $.~\QED

From propositions~\ref{Prop4} and~\ref{Prop5} we infer that the
requirement for the components of ${\mathcal D}^{\gamma }$ to vanish
along  a  path $\gamma$ determines  the corresponding normal bases with
some arbitrariness only  on $\gamma (J)$ and leaves them absolutely
arbitrary outside  the  set
$\gamma (J)$.  For this reason we speak about normal bases for
${\mathcal D}^{\gamma }$ defined only \emph{on} $\gamma (J)$.

	\begin{Prop}	\label{Prop6}
 Let the basis $\{E_{i}^{\prime }\}$ defined on $\gamma (J)$ be normal
for some S-derivation ${\mathcal D}^{\gamma }$ along a $C^{1}$ path
$\gamma :J\to M$.  Let $U$ be a coordinate neighborhood  such  that in
$U\bigcap (\gamma (J))\neq\emptyset $  the path $\gamma $ is without
self-intersections.
Then there  is a neighborhood of $U\bigcap (\gamma (J))$ in $U$ in
which $\{E_{i}^{\prime }\}$ can be extended  to
a coordinate basis, i.e. in this neighborhood there exist local
coordinates $\{y^{i }\}$ such that
$\left. E_{i}^{\prime }\right|_{\gamma (s)}=
\left. \partial/\partial y^{i }\right|_{\gamma (s)}$.
          \end{Prop}

\textbf{Remark 1.} This  proposition  means  that
locally any
normal basis for ${\mathcal D}^{\gamma }$  on $\gamma (J)$  can be
thought of as (extended to) a coordinate, and hence holonomic, one (see
proposition~\ref{Prop5}).  In particular, if $\gamma $ is contained in
only one coordinate neighborhood and is without self-intersections,
then  every normal frame on $\gamma (J)$  for ${\mathcal D}^{\gamma }$
can be extended to a holonomic one (see the proof of
proposition~\ref{Prop5}).

\textbf{Remark 2.} This result is independent of the  torsion  of  the
 derivation $D$ which induces ${\mathcal D}^{\gamma }$.
The cause for this is the condition
$\left.X\right|_{\gamma (s)}=\dot{\gamma }(s)$ in~(\ref{16}).

\textit{Proof.} The proposition is a trivial corollary from the  proof
of proposition~\ref{Prop4} and the following lemma.

	\begin{Lem}	\label{Lem7}
 Let the path $\gamma :J\to M$  be  without  self-intersections
 and such that $\gamma (J)$ is contained in some coordinate
neighborhood $U$, i.e. $\gamma (J)\subset U$.
Let $\{E_{i}^{\prime }\}$ be a smooth basis defined on
$\gamma (J)$, i.e.
$\left. E_{i}^{\prime }\right|_{\gamma (s)}$
depends  smoothly on~$s$. Then there is a neighborhood of $\gamma(J)$
in $U$  in  which  coordinates $\{y^{i }\}$  exist  such that
$\left. E_{i}^{\prime }\right|_{\gamma (s)}=
\left. \partial /\partial y^{i }\right|_{\gamma (s)}$,
i.e. $\{E_{i}^{\prime }\}$
can be extended in it to a coordinate basis.~\QED
           \end{Lem}

\textit{Proof of lemma~\ref{Lem7}.}
Let $\eta :J\times V\to U$, $V:=J\times \cdots \times J$ ($n-1$ times),
be a $C^{1}$ one-to-one map such that
$\eta (\cdot ,\mathbf{t}_{0})=\gamma $ for some  fixed
$\mathbf{t}_{0}\in V$, i.e. $\eta (s,\mathbf{t}_{0})=\gamma (s)$,
$s\in J$ (cf. the proof of proposition~\ref{Prop1}).
	 In  the neighborhood $\eta (J,V)\subset U$ we
introduce coordinates $\{x^{i}\}$  by  putting
$(x^{1}(\eta (s,\mathbf{t})),\ldots ,x^{n}(\eta (s,\mathbf{t})))=
(s,\mathbf{t})$, $s\in J$, $\mathbf{t}\in V$.
Let the nondegenerate matrix
$[  A^{i}_{j }(s;\gamma )  ]$
%
%
define the expansion of $\{E_{i}^{\prime }\}$ with respect to
$\{\partial /\partial x^{i}\}$, i.e.
	 \begin{equation}	\label{19}
\left. E_{j}^{\prime }\right|_{\gamma (s)} =
 A^{j}_{j }(s;\gamma )
\left(
\left. \frac{\partial}{\partial x^j} \right|_{\gamma(s)}
\right).
	\end{equation}

Define the functions $y^{i }:\eta (J,V)\to \mathbb{R}$ by
	\begin{eqnarray}	\nonumber
\! y^{i }(\eta (s,\mathbf{t}))
\!\! \!\! \! &:=& \!\! \!\! \!
x^{i}_{0}  +
\int^{s}_{s_{0}} (A^{-1})^{i }_{1}(u;\gamma )du +
(A^{-1})^{i }_{j}(s;\gamma )
[x^{j}(\eta (s,\mathbf{t}))-x^{j}(\gamma (s))] +
 \\ \label{20}
  &+& \!\!\!
f^{i }_{jk}(s,\mathbf{t};\gamma )
[x^{j}(\eta (s,\mathbf{t}))-x^{j}(\gamma (s))]
[x^{k}(\eta (s,\mathbf{t}))-x^{k}(\gamma (s))]
\ \ \ \
	\end{eqnarray}
where $s_{0}\in J$ and $x_{0}\in\eta(J,V)$ are fixed and the functions
$f^{i }_{jk}$  together with their first derivatives are bounded when
$\mathbf{t}\to \mathbf{t}_{0}$. Then,  because  of
$\eta (\cdot ,\mathbf{t}_{0})=\gamma $, we find
	\begin{equation}	\label{21}
\left. \frac{\partial y^{i } }{\partial x^j }\right|_{\gamma (s)} =
\left. \frac{\partial y^{i } }{\partial x^{j} }
\right|_{\eta (s,\mathbf{t}_{0}) } = (A^{-1})^{i}_{j }(s;\gamma ).
	\end{equation}

As $\det[A^{i}_{j }(s;\gamma )] \neq 0,\infty $, from~(\ref{21}) it
follows  that  the  transformation $\{x^{i}\}\to \{y^{i }\}$ is
a diffeomorphism on some neighborhood of $\gamma (J)$ lying in $U$.
So, in this neighborhood $\{y^{i }\}$ are local  coordinates.
The coordinate basic vectors on $\gamma (J)$
corresponding to them are (see~(\ref{21}) and~(\ref{19}))
\[
\left. {\partial \over \partial y^{j } }\right|_{\gamma (s)} =
\left( \left. {\partial x^{i}\over \partial y^{j } }
\right|_{\gamma (s)} \right)
\left. {\partial \over \partial x^{i}}\right|_{\gamma (s)} =
A^{i}_{j }(s;\gamma )
\left. {\partial \over \partial x^{i}}\right|_{\gamma (s)} =
\left. E_{j}^{\prime } \right|_{\gamma (s)}.
\]
Hence $\{y^{i }\}$ are the local coordinates we are looking for.~\QED

Lemma~\ref{Lem7} has also a separate  meaning: according to it any
 locally smooth basis defined \emph{on} $\gamma (J)$ can locally  be
extended to a  \textit{holonomic} basis \emph{in a neighborhood of}
$\gamma (J)$. Evidently, such an extension can be done in an
anholonomic way too.
Consequently, the holonomicity problem for a basis defined only on
$\gamma (J)$  depends  on the way this basis is extended in a
neighborhood of $\gamma (J)$.

\section {\bfseries Derivations along a fixed vector field}
\label{IV}
\setcounter {equation} {0}

Results, analogous to the ones of Sect.~\ref{II}, are true also for
 S-derivations $D_{X}$ along a \emph{fixed  vector  field} $X$
(see Sect.~\ref{2new}), in other words for a fixed derivation.
This case is briefly considered below.

	\begin{Prop}	\label{Prop8}
The S-derivation $D_{X}$ along
a fixed vector field $X$ is linear  along a path $\gamma :J\to M$,
i.e.~(\ref{12}) holds for that fixed $X$, iff along $\gamma $
 a normal frame $\{E_{i}^{\prime }\}$ for $D_{X}$ exists, i.e. one in
which the components of $D_{X}$ vanish on $\gamma (J)$.
	\end{Prop}

\textit{Proof.} If~(\ref{12}) is valid for the given $X$, then by the
proof of proposition~\ref{Prop1},  equation~(\ref{13})
has solutions $A$ given by~(\ref{14}). Consequently in the basis
$\{E_{j}^{\prime }=A^{i}_{j }E_{i}\}$ we have
$W_{X}^\prime(\gamma (s)) =
\left. [A^{-1}(W_{X}A+X(A))]\right|_{\gamma (s)}=
[\left. (A^{-1}X^{k}) \right|_{\gamma (s)}]
[\left. (\Gamma _{k}A + E_{k}(A))\right|_{\gamma (s)}]\equiv 0.$
Conversely,   if   in
$\{E_{j}^{\prime }=A^{i}_{j }E_{i}\}$  we  have
$W_{X}^\prime(\gamma (s))=0,$  then  due  to~(\ref{5'})
$\left. (W_{X}A+X(A))\right|_{\gamma (s)}=0$
is valid, i.e.
$W_{X}(\gamma (s))=\Gamma _{k}(\gamma (s))X^{k}(\gamma (s))$  for
$\Gamma _{k}(\gamma (s)) =
- \left. [(E_{k}(A))A^{-1}]\right|_{\gamma (s)}$
for the fixed vector field $X$.~\QED

      Evidently, infinitely many $\Gamma_k$'s can be found for
which~(\ref{12}) holds for a fixed  $X$. Consequently, for $D_X$ with a
fixed  $X$ there always exist normal frames along any path $\gamma$.
These frames will be explicitly constructed elsewhere for any subset of
$M$.

	\begin{Prop}	\label{Prop9}
The normal bases along $\gamma $ for $D_{X}$ for a fixed $X$ are
connected by linear transformations whose matrices are such that  the
action of $X$ on  them vanishes on $\gamma (J)$.
	\end{Prop}

\textit{Proof.} If in $\{E_{i}\}$  and
$\{E_{j}^{\prime }=A^{i}_{j }E_{i}\}$  we have
$W_{X}(\gamma (s))=W_{X}^\prime(\gamma (s))=0,$ then due to~(\ref{5'})
$\left.X(A)\right|_{\gamma (s)}=0$
is valid with
$A:=[  A^{i}_{j }  ] $, i.e.
$\left.X(A)\right|_{\gamma (J)}=0.$~\QED

For a fixed vector field $X$ the analogue of
proposition~\ref{Prop3} is, generally, not true. But if for $D_{X}$,
$X$ being fixed,~(\ref{16}) is valid on $\gamma (J)$, then we can
construct a class of S-derivations $\{{^\prime\!} D\}$ whose components
for every $X$ are given by~(\ref{12}).  Evidently,  for these
derivations proposition~\ref{Prop3} holds. Thus we have proved

	\begin{Prop}	\label{Prop10}
If along $\gamma $ for $D_{X}$ with a fixed $X$~(\ref{12})  is  valid
and there is a local holonomic (on $\gamma (J)$) normal frame along
$\gamma$ for $D_{X}$, then the above described
derivations $\{ {^\prime\!}D \}$ are torsion free on $\gamma (J)$.
Conversely, if $\{ {^\prime\!}D \}$  are  torsion
free on $\gamma (J)$ and there exists a  smooth normal frame
for $D_{X}$, then between them exist  holonomic  ones,
but generally not all of them are such.
    \end{Prop}

\section {\bfseries The case of linear connections}
\label{V}
\setcounter {equation} {0}

In this section we apply the preceding results about normal frames
to the special case of a \emph{linear connection} $\nabla $.

	\begin{Cor}	\label{Cor11}
For any linear connection  $\nabla $ there exists along every path
$\gamma:J\to M$  a field of bases in which the components of $\nabla $
vanish  on $\gamma (J)$.
These bases are  connected  with  one another  in  the  way
described in proposition~\ref{Prop2}.
	\end{Cor}

\textit{Proof.} This result is a consequence from~(\ref{7}),
propositions~\ref{Prop1} and~\ref{Prop2} and their proofs; in the
former of the proofs a basis with  the needed property is explicitly
constructed.~\QED

	\begin{Cor}	\label{Cor12}
One, and hence any, basis for a linear  connection $\nabla$ which is
smooth on $\gamma (J)$  and normal  along  a path $\gamma :J\to M$,
is holonomic if and  only if $\nabla $ is torsion free on $\gamma(J)$.
	\end{Cor}

\textbf{Remark.} If $\gamma $ is without self-intersections and
$\gamma(J)$
lies in only one coordinate neighborhood, then there exist
holonomic  normal bases (coordinates)  for $\nabla$  on $\gamma (J)$
if $\nabla $  is \emph{torsion free} and vice versa, which is a well
known fact~\cite{K&N,Schouten/Ricci,Rashevskii,Lovelock-Rund}.

\textit{Proof.} The statement follows from~(\ref{7}) and
propositions~\ref{Prop1} and~\ref{Prop3}.~\QED

	\begin{Cor}	\label{Cor13}
Let $\nabla $ be a torsion-free linear  connection  and the path
$\gamma :J\to M$ be without self-intersections and lying in only one
coordinate neighborhood. Then for  $\nabla$ there exist
normal coordinates on  $\gamma (J)$, or, equivalently, locally
holonomic normal bases.
	\end{Cor}

\textbf{Remark.} This corollary reproduces a  classical  theorem  that
can be found, for instance, in~\cite{Rashevskii} or
in~\cite[ch.  III,  \S 8]{Schouten/Ricci}, in the latter of which
references to original papers are given.

\textit{Proof.} The result follows from corollaries~\ref{Cor11}
and~\ref{Cor12}.~\QED

	\begin{Cor}	\label{Cor14}
Let
$\left.\frac{D}{ds}\right|_{\gamma }:=\nabla _{\dot{\gamma }}$
be the covariant derivative associated with $\nabla $  along the
$C^{1}$ path $\gamma:J\to M$. Then on $\gamma (J)$ there exist normal
frames for $\nabla _{\dot{\gamma }}$.
They are obtained from one another by linear
transformations whose  coefficients are constant or depend only on
$\gamma $. If $\gamma $ is without self-intersections and $\gamma (J)$
lies in only one coordinate neighborhood, then in  some neighborhood
of $\gamma (J)$ all of
these normal frames can be extended in  a holonomic way.
	\end{Cor}

\textit{Proof.} The statement follows from propositions~\ref{Prop4},
\ref{Prop5}, and~\ref{Prop6}.~\QED

\section {\bfseries Conclusion}
\label{VI}
\setcounter {equation} {0}

The  above  investigation  shows  that  under  sufficiently general
conditions there exist, generally anholonomic, bases in which
the components of a derivation of the tensor algebra
over a differentiable manifold $M$ vanish
along a path $\gamma :J\to M$. These bases (frames) are called
\emph{normal}. When the derivations are along paths, then the
corresponding normal bases always can be taken as holonomic
(or coordinate) ones. These results generalize a series of analogous
ones concerning linear connections and originating from~\cite{Fermi}.

A feature of the  case  along  paths considered here  is  its
independence of the derivation's  curvature,  which wasn't even
introduced here. The cause for this is the one  dimensionality  of
the paths (curves) considered as submanifolds of $M$. In  this
connection it is interesting to consider the analogous problems
on arbitrary submanifolds of $M$,  which will be done elsewhere.

Now we shall consider  briefly the relation of the results obtained in
this paper with the equivalence principle~\cite{MTW,Heyde}.
According
to it the gravitational field strength, usually identified with
the components of some linear connection, is transformable
to zero at a point by an appropriate choice of the  local
(called normal, geodesic, Riemannian, inertial, or  Lorentz)
coordinates  or  reference  frame
(basis). So, from a mathematical  point  of  view,  the  equivalence
principle states the existence of local bases in which the
corresponding connection's components vanish at a point. The results of
this investigation show the strict validity of this statement along any
path (curve). Hence, we can make the following  three  conclusions:
(i) Any gravitational theory based on space-time  with  a
linear connection is compatible  with  the  equivalence  principle
along every path, i.e. in it there exist (local) inertial frames along
paths. These frames are generally anholonomic, but under some
(not very restrictive from a physical point of view) conditions on the
paths (see lemma~\ref{Lem7}) there exist such  holonomic  frames  of
reference.
(ii) In gravitational theories based on  linear
connections the equivalence principle along paths must not be
considered as a principle (in a sense of an axiom) as it is
identically fulfilled because of their mathematical background.
(iii) If we want the equivalence
principle along paths  to be valid in gravitational theories based on
some (class of) tensor derivations
(cf.~\cite[Sect.V]{f-Bases-n+point}),
then this principle will select only  the theories based on linear
connections, i.e.  only those in which  it is identically satisfied.
In fact, suppose the gravitational field strength to be locally
identified with the components of a certain tensor derivation. The
equivalence principle along paths requires the vanishment of the
gravitational field strength along paths. So, this leads to the
possibility to transform the components of the tensor derivation to
zero along \emph{any} path. By proposition~\ref{Prop1} this implies the
derivation to be linear along every path which is possible iff it is
linear at every point, i.e. iff it is a linear connection.

\section * {\bfseries Acknowledgements}
\label{VII}
\setcounter {equation} {0}

	The author expresses his gratitude to Dr. Sawa Manoff
(Institute for Nuclear  Research  and  Nuclear Energy
of the Bulgarian Academy of Sciences)
for valuable comments and stimulating discussions.

	This work was partially supported by
the National Foundation for Scientific Research of Bulgaria
under Grant No. F642.



\begin{thebibliography}{10}

\bibitem{K&N}
Kobayashi S. and Nomizu K.
\newblock {\em Foundations of Differential Geometry}, volume~I.
\newblock Interscience Publishers, New York-London, 1963.

\bibitem{Schouten/Ricci}
Schouten~J. A.
\newblock {\em Ricci-Calculus: An Introduction to Tensor Analysis and
its   Geometrical Applications}.
\newblock Springer Verlag, Berlin-G{\"o}ttingen-Heidelberg, 1954.

\bibitem{Rashevskii}
Rashevskii~P. K.
\newblock {\em Riemannian Geometry and Tensor Analysis}.
\newblock Nauka, Moscow, 1967.
\newblock (In Russian).

\bibitem{Fermi}
Fermi E.
\newblock Sopra i fenomeni che avvengono in vicinonza di una linear
oraria  ({A}bout phenomenons near a world line).
\newblock {\em Atti {R}. {A}ccad {L}incei {R}end., {C}l. {S}ci. {F}is.
{M}at. {N}at.}, 31(1):21--23, 51--52, 101--103, 1922.

\bibitem{Eisenhart}
Eisenhart~L. P.
\newblock {\em Non-Riemannian Geometry}.
\newblock Am. Math. Soc. Cal. Publ., New York, 1927.

\bibitem{Heyde}
von~der Heyde~P.
\newblock The equivalence principle in the ${U}_4$ theory of
gravitation.
\newblock {\em Lettare al {N}uovo {C}imento}, 14(7):250--252, 1975.

\bibitem{MTW}
Misner~C. W., Thorne~K. S., and Wheeler~J. A.
\newblock {\em Gravitation}.
\newblock W. H. Freeman and Company, San Francisco, 1973.

\bibitem{f-Frames-n+point}
Iliev~B. Z.
\newblock Normal frames and the validity of the equivalence principle:
I. {C}ases in a neighborhood and at a point.
\newblock {\em J. Physics A: Math. and Gen.}, 29(21):6895--6901, 1996.
\newblock (HEP Preprint Server \mbox{No.~gr-qc/9608019}).

\bibitem{f-Bases-path}
Iliev~B. Z.
\newblock Special bases for derivations of tensor algebras.
{II}.~{C}ase along   paths.
\newblock JINR Communication E5-92-508, Dubna, 1992.

\bibitem{f-Bases-n+point}
Iliev~B. Z.
\newblock Special bases for derivations of tensor algebras.
{I}.~{C}ases in a   neighborhood and at a point.
\newblock JINR Communication E5-92-507, Dubna, 1992.

\bibitem{Lovelock-Rund}
Lovelock D. and Rund H.
\newblock {\em Tensors, Differential Forms, and Variational
Principals}.
\newblock Wiley-Interscience Publication, John Wiley {\&} Sons, New
  York-London-Sydney-Toronto, 1975.

\bibitem{Hartman}
Hartman Ph.
\newblock {\em Ordinary Differential Equations}.
\newblock John Wiley {\&} Sons, New York-London-Sydney, 1964.

\end{thebibliography}

\end{document}